\documentclass[prd,email,twocolumn,showpacs,showkeys,preprintnumbers,amsmath,amssymb,nofootinbib]{revtex4-1}
\usepackage{amsmath}
\usepackage{latexsym}

\begin{document}

\title{Coulomb's law corrections from a gauge-kinetic mixing}

\author{Patricio Gaete}
\email{patricio.gaete@usm.cl}
\author{Iv\'an  Schmidt}
\email{ivan.schmidt@usm.cl}

\affiliation{${}$Departmento de F\'{\i}sica and Centro
Cient\'{\i}fico-Tecnol\'{o}gico
de Valpara\'{\i}so, \\
Universidad T\'ecnica Federico Santa Mar\'{\i}a, Valpara\'{\i}so, Chile \\
\today\\}



\begin{abstract}
\noindent We study the static quantum potential for a gauge theory
which includes the mixing between the familiar photon $U(1)_{QED}$
and a second massive gauge field living in the so-called
hidden-sector $U(1)_h$. Our discussion is carried out using the
gauge-invariant but path-dependent variables formalism, which is
alternative to the Wilson loop approach. Our results show that the
static potential is a Yukawa correction to the usual static Coulomb
potential. Interestingly, when this calculation is done inside a
superconducting box, the Coulombic piece disappears leading to a
screening phase.
\end{abstract}

\maketitle

\pagestyle{myheadings} \markright{{\it Coulomb's law corrections
from a gauge-kinetic mixing}}

The formulation and experimental consequences of extensions of the
Standard Model (SM), such as axion-like particles and light extra
hidden $U(1)$ gauge bosons have been extensively discussed in the
last few years, in order to explain cosmological and astrophysical
results \cite{Zavattini}. In this case, hidden sector refers to a
set of so far unobserved degrees of freedom very weakly coupled  to
the SM. We further note that the existence of weakly interacting
particles, which couple to the electromagnetic sector, also are of
interest due to their appearance in string theories physics.
Nevertheless, although none of these searches ultimately has yielded
a positive signal, the arguments in favor of the existence of
axion-like particles or light extra hidden $U(1)$ gauge bosons,
remain a relevant and challenging issue.

It is worthy recalling at this stage that the axion-like scenario
can be qualitatively understood by the existence of light
pseudo-scalar bosons $\phi$ (``axions''), with a coupling to two
photons. As a consequence, the corresponding term in the effective
Lagrangian has the form $ {\cal L}_I  = - {\raise0.7ex\hbox{$1$}
\!\mathord{\left/ {\vphantom {1
{4M}}}\right.\kern-\nulldelimiterspace}
\!\lower0.7ex\hbox{${4M}$}}F_{\mu \nu } {\widetilde F}^{\mu \nu}
\phi$, where $ {\widetilde F}^{\mu \nu }  = {\raise0.7ex\hbox{$1$}
\!\mathord{\left/{\vphantom {1 2}}\right.\kern-\nulldelimiterspace}
\!\lower0.7ex\hbox{$2$}}\varepsilon _{\mu \nu \lambda \rho }
F^{\lambda \rho }$. We also recall that axionic electrodynamics
experiences mass generation due to the breaking of rotational
invariance induced by a classical background configuration of the
gauge field strength \cite{Spallucci}, and leads to confining
potentials in the presence of nontrivial constant expectation values
for the gauge field strength $F_{\mu \nu}$ \cite{GaeteGuen}. In
fact, in the case of a constant electric field strength expectation
value, the static potential remains Coulombic, while in the case of
a constant magnetic field strength expectation value the potential
energy is the sum of a Yukawa and a linear potential, leading to the
confinement of static charges. Another interesting observation is
that the magnetic character of the field strength expectation value
needed to obtain confinement is in agreement  with the current
chromo-magnetic picture of the $QCD$ vacuum
\cite{Savvidy,Ninomiya,Nielsen}. In addition, similar results have
been obtained in the context of the dual Ginzburg-Landau theory
\cite{Suganuma}, as well as for a theory of antisymmetric tensor
fields that results from the condensation of topological defects as
a consequence of the Julia-Toulouse mechanism \cite{GaeteW}.

As already expressed, the extra hidden $U(1)$ gauge bosons scenario,
which  is described by the mixing between the familiar photon
$U(1)_{QED}$ and a second gauge field (paraphoton) living in the
hidden-sector $U(1)_h$, has been inspired by studies coming from the
realms of string theory \cite{Jaeckel2,Batell}, as well as from
quantum field theory
\cite{Holdom,Nath,Nath2,Castelo,Masso,Foot,Singleton,Langacker}.
Actually, as mentioned in \cite{Singleton2}, the introduction of a
second gauge field in addition to the usual photon was pioneered in
Ref. \cite{Cabibbo}, in the context of electrodynamics in the
presence of magnetic monopoles \cite{Dirac}. The quantization for a
system with two photons was later carried out in \cite{Hagen}. Let
us also mention here that the possible existence of massive vector
fields was also proposed in \cite{Okun}. Most recently, a kinetic
term between the familiar photon $U(1)_{QED}$ and a second gauge
field has been considered in order to explain recent unexpected
observations in high energy astrophysics \cite{Nima}.

Given the ongoing experiments related to this type of physics
\cite{Ehret,Afanasev,Pugnat,Cantatore}, it should be interesting to
acquire a better understanding what might be the observational
signatures presented by the hidden-sector $U(1)_h$. Hence, our
purpose here is to investigate the impact of paraphotons on physical
observables, in particular the static potential between two charges,
using the gauge-invariant but path-dependent variables formalism,
which is alternative to the Wilson loop approach.  Of special
interest will be to explore  the existence of duality  (equivalence)
between  axion-like particles and light extra hidden $U(1)$ gauge
bosons models, by comparing physical quantities in both theories.

 We now proceed to analyze the interaction energy between static
point-like sources for the model under consideration. The main tool
in our analysis will be to compute the expectation value of the
energy operator $H$ in the physical state $|\Phi\rangle$ describing
the sources, which we will denote by $ {\langle H\rangle}_\Phi$.

As we have already pointed out, the gauge theory we are considering
describes the coupling between the familiar massless
electromagnetism $U(1)_{QED}$ and a hidden-sector $U(1)_h$. In this
case the corresponding theory is governed by the Lagrangian density
\cite{Ringwald}:
\begin{equation}
{\cal L} =  - \frac{1}{4}F_{\mu \nu } F^{\mu \nu }  -
\frac{1}{4}B_{\mu \nu } B^{\mu \nu }  - \frac{1}{2}\chi F^{\mu \nu }
B_{\mu \nu } + \frac{1}{2}m_{\gamma ^ \prime  }^2 B_\mu  B^\mu,
\label{Pho5}
\end{equation}
where $\chi$ is a dimensionless parameter. Recalling again that the
theory we are considering emerges naturally in extensions of the
Standard Model in the context of string theories with nonvanishing
values for $\chi$ \cite{Jaeckel2,Dienes}. We also note here that in
expression (\ref{Pho5}) the first two terms are the standard kinetic
terms for the photon and hidden-sector photon fields, respectively.
The third term is the so called gauge-kinetic mixing, while the last
term accounts for a possible mass of the paraphoton, which may arise
via a Higgs or St\"{u}eckelberg mechanism. After shifting the
$B_\mu$-field as ${B}_\mu\longrightarrow \tilde{B}_\mu-\chi A_\mu$
in (\ref{Pho5}), we obtain an equivalent Lagrangian density:
\begin{eqnarray}
{\cal L} &=&  - \frac{1}{4}\left( {1 - \chi ^2 } \right)F_{\mu \nu }
F^{\mu \nu }  + \frac{1}{2}\chi ^2 m_{\gamma ^ \prime  }^2 A_\mu
A^\mu \nonumber \\
&+& \frac{1}{2}\tilde{B}^\mu  \left[ {\eta _{\mu \nu } \left(
{\Delta + m_{\gamma ^ \prime  }^2 } \right) - \partial _\mu
\partial _\nu } \right]\tilde{B}^\nu \nonumber \\
&-& \chi m_{\gamma ^ \prime  }^2 A_\mu \tilde{B}^\mu. \label{Pho10}
\end{eqnarray}
Next, by integrating out the $\tilde{B}_\mu$-field in the above
expression, one gets an effective action for the $A_\mu$-field that
includes the effects of the hidden-sector photon field:
\begin{equation}
 {\cal L}_{eff} =  - \frac{1}{4}F_{\mu \nu } \left( {1 - \chi ^2
\frac{\Delta }{{\left( {\Delta  + m_{\gamma ^ \prime  }^2 }
\right)}}} \right)F^{\mu \nu }. \label{Pho15}
\end{equation}
Let us also mention here that in addition to the above
transformation to diagonalize the kinetic mixing term, we can also
choose the transformation ${A}_\mu\longrightarrow \tilde{A}_\mu-\chi
B_\mu$, $\tilde{B}_\mu = B_\mu$. As a consequence, there is no
kinetic mixing between $\tilde{A}_\mu$ and $\tilde{B}_\mu$, and one
obtains two noninteracting gauge theories with a trivial vacuum. In
this perspective, the transformation leading to (\ref{Pho15})
induces a richer dynamics that incorporate the effects of the
hidden-sector photon field. In other words, this analysis renders
manifest the signatures of the hidden-sector $U(1)_h$ on the
ordinary photon sector.

Having characterized the theory under study, we can now compute the
interaction energy. To this end, we shall first examine the
Hamiltonian framework for this theory. The canonical momenta $\Pi
^\mu$, conjugate to $A ^\mu$, are found to be $\Pi^{0}=0$, which is
the primary constraint, while the other momenta components are $\Pi
^i = - \frac{{\left( {1 - \chi ^2 }\right)\Delta + m_{\gamma ^
\prime }^2 }}{{\left( {\Delta + m_{\gamma ^ \prime }^2 }
\right)}}F^{0i}$. The canonical Hamiltonian is now obtained in the
usual way
\begin{equation}
H_C = \int {d^3 x} \left\{ { - A^0 \partial _i \Pi ^i  +
\frac{1}{{2\left( {1 - \chi ^2 } \right)}}\Pi ^i \frac{{\left(
{\Delta  + m_{\gamma ^ \prime  }^2 } \right)}}{{\left( {\Delta  +
M_{\gamma ^ \prime  }^2 } \right)}}\Pi ^i } \right\} \nonumber \\
\end{equation}
\begin{equation}
+ \int {d^3 x} \left\{ {\frac{{\left( {1 - \chi ^2 }
\right)}}{4}B^i \frac{{\left( {\Delta  + M_{\gamma ^ \prime  }^2 }
\right)}}{{\left( {\Delta  + m_{\gamma ^ \prime  }^2 } \right)}}B^i
} \right\}, \label{Pho20}
\end{equation}
where $M_{\gamma ^ \prime  }^2  \equiv {\raise0.7ex\hbox{${m_{\gamma
^ \prime  }^2 }$} \!\mathord{\left/
 {\vphantom {{m_{\gamma ^ \prime  }^2 } {\left( {1 - \chi ^2 }
\right)}}}\right.\kern-\nulldelimiterspace}
\!\lower0.7ex\hbox{${\left( {1 - \chi ^2 } \right)}$}}$, and $B^i$
is the magnetic field. Persistency of the primary constraint in time
yields a secondary constraint, which is the usual Gauss constraint
$\Gamma_1 \left( x\right) \equiv \partial _i \Pi ^i=0$, and time
stability of this secondary constraint does not induce further
constraints. Therefore, the extended Hamiltonian that generates
translations in time reads $H = H_C + \int {d^3 }x\left( {c_0 \left(
x \right)\Pi _0 \left( x \right) + c_1 \left( x\right)\Gamma _1
\left( x \right)} \right)$. Here $c_0 \left( x\right)$ and $c_1
\left( x \right)$ are arbitrary Lagrange multipliers. Moreover, it
is straightforward to see that $\dot{A}_0 \left( x \right)= \left[
{A_0 \left( x \right),H} \right] = c_0 \left( x \right)$, which is
an arbitrary function. Since $ \Pi^0 = 0$ always, neither $ A^0 $
nor $ \Pi^0 $ are of interest in describing the system and may be
discarded from the theory. If a new arbitrary coefficient $c(x) =
c_1 (x) - A_0 (x)$ is introduced the Hamiltonian may be rewritten in
the form
\begin{eqnarray}
H &=& \int {d^3 x} \left\{ { c(x) \partial _i \Pi ^i  +
\frac{1}{{2\left( {1 - \chi ^2 } \right)}}\Pi ^i \frac{{\left(
{\Delta  + m_{\gamma ^ \prime  }^2 } \right)}}{{\left( {\Delta  +
M_{\gamma ^ \prime  }^2 } \right)}}\Pi ^i } \right\} \nonumber \\
&+& \int {d^3 x} \left\{ {\frac{{\left( {1 - \chi ^2 }
\right)}}{4}B^i \frac{{\left( {\Delta  + M_{\gamma ^ \prime  }^2 }
\right)}}{{\left( {\Delta  + m_{\gamma ^ \prime  }^2 } \right)}}B^i
} \right\}. \label{Pho25}
\end{eqnarray}

In accordance with the Dirac method, we must fix the gauge. A
particularly convenient gauge fixing condition is
\begin{equation}
\Gamma _2 \left( x \right) \equiv \int\limits_{C_{\xi x} } {dz^\nu }
A_\nu \left( z \right) \equiv \int\limits_0^1 {d\lambda x^i } A_i
\left( {\lambda x} \right) = 0, \label{Pho30}
\end{equation}
where  $\lambda$ $(0\leq \lambda\leq1)$ is the parameter describing
the spacelike straight path $ x^i = \xi ^i  + \lambda \left( {x -
\xi } \right)^i $, and $ \xi $ is a fixed point (reference point).
There is no essential loss of generality if we restrict our
considerations to $ \xi ^i=0 $. The choice (\ref{Pho30}) leads to
the Poincar\'e gauge \cite{Pato,GaeteSprd}. With this, we arrive at
the only nonvanishing equal-time Dirac bracket for the canonical
variables
\begin{eqnarray}
\left\{ {A_i \left( x \right),\Pi ^j \left( y \right)} \right\}^ *
&=&\delta{ _i^j} \delta ^{\left( 3 \right)} \left( {x - y}
\right)\nonumber \\
&-& \partial _i^x \int\limits_0^1 {d\lambda x^j } \delta ^{\left( 3
\right)} \left( {\lambda x - y} \right). \label{Pho35}
\end{eqnarray}

Having thus obtained the quantization of the model under
consideration, we are now in a position to compute the interaction
energy. As already expressed, to this end we will work out the
expectation value of the energy operator $H$ in the physical state
$|\Phi\rangle$, which has the important property of being
gauge-invariant \cite{Dirac2,Dirac3}, and has the stringy form
\begin{eqnarray}
\left| \Phi  \right\rangle  &\equiv& \left| {\overline \Psi  \left(
\bf y \right)\Psi \left( {\bf y}\prime \right)} \right\rangle
\nonumber \\
&=& \overline \psi \left( \bf y \right)\exp \left(
{iq\int\limits_{{\bf y}\prime}^{\bf y} {dz^i } A_i \left( z \right)}
\right)\psi \left({\bf y}\prime \right)\left| 0 \right\rangle,
\label{Pho40}
\end{eqnarray}
where $\left| 0 \right\rangle$ is the physical vacuum state and the
line integral appearing in the above expression is along a spacelike
path starting at ${\bf y}\prime$ and ending at $\bf y$, on a fixed
time slice.

Taking into account the above Hamiltonian structure, we observe that
\begin{eqnarray}
\Pi _i \left( x \right)\left| {\overline \Psi  \left( \bf y
\right)\Psi \left( {{\bf y}^ \prime  } \right)} \right\rangle  &=&
\overline \Psi  \left( \bf y \right)\Psi \left( {{\bf y}^ \prime }
\right)\Pi _i \left( x \right)\left| 0 \right\rangle \nonumber\\
&+&  q\int_ {\bf y}^{{\bf y}^ \prime  } {dz_i \delta ^{\left( 3
\right)} \left( {\bf
z - \bf x} \right)} \left| \Phi \right\rangle. \nonumber\\
\label{Pho45}
\end{eqnarray}
Having made this observation and since the fermions are taken to be
infinitely massive (static) we can substitute $\Delta$ by
$-\nabla^{2}$ in Eq. (\ref{Pho25}). Therefore, the expectation value
$\left\langle H \right\rangle _\Phi$ is expressed as
\begin{equation}
\left\langle H \right\rangle _\Phi   = \left\langle H \right\rangle
_0 + \left\langle H \right\rangle _\Phi ^{\left( 1 \right)}  +
\left\langle H \right\rangle _\Phi ^{\left( 2 \right)},
\label{Pho50}
\end{equation}
where $\left\langle H \right\rangle _0  = \left\langle 0
\right|H\left| 0 \right\rangle$. The $\left\langle H \right\rangle
_\Phi ^{\left( 1 \right)}$ and $\left\langle H \right\rangle _\Phi
^{\left( 2 \right)}$ terms are given by
\begin{equation}
\left\langle H \right\rangle _\Phi ^{\left( 1 \right)}  =  -
\frac{1}{2} \left\langle \Phi  \right|\int {d^3 x} \Pi _i \Pi ^i
\left| \Phi  \right\rangle, \label{Pho50a}
\end{equation}
and
\begin{equation}
\left\langle H \right\rangle _\Phi ^{\left( 2 \right)}  =  -
\frac{1}{2} \frac{{\chi ^2}} {{(1 - \chi ^2) }}   \left\langle \Phi
\right|\int {d^3 x} \Pi_i \frac{{\nabla ^2 }}{{\left( {\nabla ^2  -
M_{\gamma ^ \prime }^2 } \right)}} \Pi^i \left| \Phi
\right\rangle.\label{Pho50b}
\end{equation}

Following our earlier procedure \cite{Gaete99,Gaete07}, we see that
the potential for two opposite charges, located at ${\bf y}$ and
${\bf y^{\prime}}$, takes the form
\begin{equation}
V =  - \frac{{q^2 }}{{4\pi }}\left( {\frac{1}{L} + \frac{{\chi ^2
}}{{(1 - \chi ^2) }} \frac{{e^{ - M_\gamma \prime L} }}{L}} \right),
\label{Pho55}
\end{equation}
where $|{\bf y} -{\bf y}^{\prime}|\equiv L$. We stress that the
choice of the gauge is in this approach really arbitrary. Being the
formalism completely gauge invariant, we would obtain exactly the
same result in any gauge.

It is worth noting here that there is an alternative but equivalent
way of obtaining the result (\ref{Pho55}), which highlights certain
distinctive features of our methodology. We start by considering
\cite{Gaete99}:
\begin{equation}
V \equiv q\left( {{\cal A}_0 \left( {\bf 0} \right) - {\cal A}_0
\left( {\bf y} \right)} \right), \label{Pho60}
\end{equation}
where the physical scalar potential is given by
\begin{equation}
{\cal A}_0 \left( {x^0 ,{\bf x}} \right) = \int_0^1 {d\lambda } x^i
E_i \left( {\lambda {\bf x}} \right), \label{Pho65}
\end{equation}
with $i=1,2,3$. This follows from the vector gauge-invariant field
expression \cite{Pato}
\begin{equation}
{\cal A}_\mu  \left( x \right) \equiv A_\mu  \left( x \right) +
\partial _\mu  \left( { - \int_\xi ^x {dz^\mu  } A_\mu  \left( z
\right)} \right), \label{Pho70}
\end{equation}
where, as in Eq.(\ref{Pho30}), the line integral is along a
spacelike path from the point $\xi$ to $x$, on a fixed slice time.
The gauge-invariant variables (\ref{Pho70}) commute with the sole
first constraint (Gauss' law), confirming that these fields are
physical variables \cite{Dirac}. Note that Gauss' law for the
present theory reads $\partial _i \Pi ^i  = J^0$, where we have
included the external current $J^0$ to represent the presence of two
opposite charges. For $J^0 \left( {t,{\bf x}} \right) = q\delta
^{\left( 3 \right)} \left( {\bf x} \right)$ the electric field is
given by
\begin{equation}
E^i  = q\partial ^i \left( {G\left( {\bf x} \right) + \frac{{\chi ^2
}} {{\left( {1 - \chi ^2 } \right)}}\tilde G\left( {\bf x} \right)}
\right), \label{Pho75}
\end{equation}
where $G\left( {\bf x} \right) = \frac{1}{{4\pi }}\frac{1}{{|{\bf x}|}}$ and
  $\tilde G\left( {\bf x} \right) = \frac{{e^{ - M_{\gamma ^ \prime  }
\left| {\bf x} \right|} }}{{4\pi \left| {\bf x} \right|}}$ are the
Green functions in three space dimensions. Finally, replacing this
result in (\ref{Pho65}) and using (\ref{Pho60}), the potential for a pair of
point-like opposite charges q located at ${\bf 0}$ and ${\bf L}$ becomes
\begin{equation}
V =  - \frac{{q^2 }}{{4\pi }}\left( {\frac{1}{L} + \frac{{\chi ^2
}}{{(1 - \chi ^2) }} \frac{{e^{ - M_\gamma \prime L} }}{L}} \right),
\label{Pho80}
\end{equation}
where $\left| {\bf L} \right| \equiv L$.

Mention should be made, at this point, about a new and simple
experiment for searching for hidden-sector $U(1)$ gauge bosons with
gauge kinetic mixing with the ordinary photon \cite{Jaeckel}.  The
crucial idea underlying this suggestion consists in putting a
sensitive magnetometer inside a superconducting shielding, which in
turn is placed inside a strong magnetic field. As a result it was
argued that photon - hidden photon oscillations would allow to
penetrate the superconductor and a magnetic field would register on
the magnetometer, in contrast with the usual electrodynamics where
the magnetic field cannot penetrate the superconductor. For this
purpose the authors of Ref. \cite{Jaeckel} consider the Lagrangian
density:
\begin{eqnarray}
{\cal L} &=&  - \frac{1}{4}F_{\mu \nu } F^{\mu \nu }  -
\frac{1}{4}B_{\mu \nu } B^{\mu \nu }  - \frac{1}{2}\chi F^{\mu \nu }
B_{\mu \nu } +\frac{1}{2}m_{\gamma ^ \prime  }^2 B_\mu  B^\mu
\nonumber\\
&+& \frac{1}{2}M_{Lon}^2 A_\mu A^\mu, \label{Pho85}
\end{eqnarray}
where the last term corresponds to the London mass ($M_{Lon}$) that
the photon acquires inside the superconductor. Notice that in vacuum
$M_{Lon}=0$, and expression  (\ref{Pho85}) reduces to expression
(\ref{Pho5}). Given its relevance, it is of interest to study the
stability of the above scenario (\ref{Pho85}) for the $M_{Lon}\neq0$
case.

Following the same steps that lead to (\ref{Pho50}) we obtain
\begin{eqnarray}
\left\langle H \right\rangle _\Phi ^{\left( 1 \right)}  &=&
\frac{1}{2}\frac{1}{{\left( {1 - \chi ^2 } \right)}}\frac{{\left( {M_1^2
- m_{\gamma ^ \prime  }^2 } \right)}}{{\left( {M_2^2  - M_1^2 } \right)}} \nonumber \\
&\times& \left\langle \Phi  \right|\int {d^3 x} \Pi _i \frac{{\nabla
^2 }} {{\left( {\nabla ^2  - M_1^2 } \right)}}\Pi ^i \left| \Phi
\right\rangle , \label{Pho90a}
 \end{eqnarray}

and
\begin{eqnarray}
\left\langle H \right\rangle _\Phi ^{\left( 2 \right)}  &=& -
\frac{1}{2}\frac{1}
{{\left( {1 - \chi ^2 } \right)}}\frac{{\left( {M_2^2  - m_{\gamma ^ \prime  }^2 } \right)}}
{{\left( {M_2^2  - M_1^2 } \right)}} \nonumber \\
&\times& \left\langle \Phi  \right|\int {d^3 x} \Pi _i \frac{{\nabla
^2 }} {{\left( {\nabla ^2  - M_2^2 } \right)}}\Pi ^i \left| \Phi
\right\rangle , \label{Pho90b}
\end{eqnarray}
 where \\
 \\
$M_1^2  = \frac{{\left( {m_{\gamma ^ \prime  }^2  + M_{Lon}^2 } \right)^2 }}
{{2\left( {1 - \chi ^2 } \right)^2 }}\left[ {1 + \sqrt {1 - 4\left( {1 - \chi ^2 }
 \right)^3 \frac{{m_{\gamma ^ \prime  }^2 M_{Lon}^2 }}{{\left( {m_{\gamma ^ \prime  }^2
 + M_{Lon}^2 } \right)^4 }}} } \right]$ \\
\\
and \\
\\
 $M_2^2  = \frac{{\left( {m_{\gamma ^ \prime }^2  + M_{Lon}^2 } \right)^2 }}
{{2\left( {1 - \chi ^2 } \right)^2 }}\left[ {1 - \sqrt {1 - 4\left( {1 - \chi ^2 } \right)^3
 \frac{{m_{\gamma ^ \prime  }^2 M_{Lon}^2 }}{{\left( {m_{\gamma ^ \prime  }^2
+ M_{Lon}^2 } \right)^4 }}} } \right]$.  \\
\\
Again, following our earlier procedure \cite{Gaete99,Gaete07}, we
see that the potential for two opposite charges, located at ${\bf
y}$ and ${\bf y^{\prime}}$, takes the form
\begin{eqnarray}
V &=&  - \frac{{q^2 }}{{4\pi }}\frac{1}{{\left( {1 - \chi ^2 }
\right)\left( {M_2^2  - M_1^2 } \right)}} \nonumber \\
&\times& \left[ {\left( {M_2^2  - m_{\gamma ^ \prime  }^2 } \right)\frac{{e^{ - M_2 L} }}{L}
- \left( {M_1^2  - m_{\gamma ^ \prime  }^2 } \right)\frac{{e^{ - M_1 L} }}{L}} \right].
\nonumber \\
\label{Pho95}
 \end{eqnarray}
Hence we see that for $M_{Lon}=0$, expression (\ref{Pho95}) reduces
to expression (\ref{Pho55}). Therefore, when the analysis is done
inside a superconducting box, the surprising result is that the
theory describes an exactly screening phase.

To conclude, once again we have exploited a crucial point for
understanding the physical content of gauge theories, that  is, the
identification of field degrees of freedom with observable
quantities. Our analysis reveals that the static potential profile
obtained from both the coupling between the familiar massless
electromagnetism $U(1)_{QED}$ and a hidden-sector $U(1)_h$ and
axionic electrodynamics models are quite different. This means that
the two theories are not equivalent.  As it was shown in
\cite{GaeteGuen}, axionic electrodynamics has a different structure
which is reflected in a confining piece, which is not present in the
coupling between the familiar massless electromagnetism $U(1)_{QED}$
and a hidden-sector $U(1)_h$ scenario.

This work was partially supported by Fondecyt (Chile), projects
1080260 and 1100287.

\end{document}